\documentclass[11pt,psfig,epsf,bbox]{article} 

\usepackage{amsmath,amssymb,amsthm}
\DeclareMathAlphabet{\mathpzc}{OT1}{pzc}{m}{it}

\begin{document}

\newcommand{\vAi}{{\cal A}_{i_1\cdots i_n}} \newcommand{\vAim}{{\cal
A}_{i_1\cdots i_{n-1}}} \newcommand{\vAbi}{\bar{\cal A}^{i_1\cdots i_n}}
\newcommand{\vAbim}{\bar{\cal A}^{i_1\cdots i_{n-1}}}
\newcommand{\htS}{\hat{S}} \newcommand{\htR}{\hat{R}}
\newcommand{\htB}{\hat{B}} \newcommand{\htD}{\hat{D}}
\newcommand{\htV}{\hat{V}} \newcommand{\cT}{{\cal T}} \newcommand{\cM}{{\cal
M}} \newcommand{\cMs}{{\cal M}^*}
 \newcommand{\vk}{{\bf k}}
\newcommand{\vK}{{\vec K}} \newcommand{\vb}{{\vec b}} \newcommand{{\vp}}{{\vec
p}} \newcommand{{\vq}}{{\vec q}} \newcommand{\vQ}{{\vec Q}}
\newcommand{\vx}{{\vec x}}
\newcommand{\tr}{{{\rm Tr}}} 
\newcommand{\beq}{\begin{equation}}
\newcommand{\eeq}[1]{\label{#1} \end{equation}} 
\newcommand{\half}{{\textstyle
\frac{1}{2}}} \newcommand{\gton}{\stackrel{>}{\sim}}
\newcommand{\lton}{\mathrel{\lower.9ex \hbox{$\stackrel{\displaystyle
<}{\sim}$}}} \newcommand{\ee}{\end{equation}}
\newcommand{\ben}{\begin{enumerate}} \newcommand{\een}{\end{enumerate}}
\newcommand{\bit}{\begin{itemize}} \newcommand{\eit}{\end{itemize}}
\newcommand{\bc}{\begin{center}} \newcommand{\ec}{\end{center}}
\newcommand{\bea}{\begin{eqnarray}} \newcommand{\eea}{\end{eqnarray}}
\newcommand{\beqar}{\begin{eqnarray}} \newcommand{\eeqar}[1]{\label{#1}
\end{eqnarray}} \newcommand{\bra}[1]{\langle {#1}|}
\newcommand{\ket}[1]{|{#1}\rangle}
\newcommand{\norm}[2]{\langle{#1}|{#2}\rangle}
\newcommand{\brac}[3]{\langle{#1}|{#2}|{#3}\rangle} \newcommand{\hilb}{{\cal
H}} \newcommand{\pleft}{\stackrel{\leftarrow}{\partial}}
\newcommand{\pright}{\stackrel{\rightarrow}{\partial}}

\begin{center}
{\Large {\bf{An overview of heavy quark energy loss puzzle at RHIC}}}

\vspace{1cm}

{ Magdalena Djordjevic}

\vspace{.8cm}

{\em { Dept. Physics, Ohio State University, 191 West Woodruff Avenue, 
Columbus, OH 43210, USA}}

\vspace{.5cm}

\today
\end{center}

\vspace{.5cm}

\begin{abstract}
We give a theoretical overview of the heavy quark tomography puzzle posed by 
recent non-photonic single electron data from central Au+Au collisions at 
$\sqrt{s} = 200$ AGeV. We show that radiative energy loss mechanisms alone are 
not able to explain large single electron suppression data, as long as 
realistic parameter values are assumed. We argue that combined collisional and 
radiative pQCD approach can solve a substantial part of the non-photonic single
electron puzzle. 
\end{abstract}

\section{Introduction}

Quark Gluon Plasma (QGP) is a new form of matter, consisting of interacting 
quarks, antiquarks and gluons. If the QGP can be created in Ultrarelativistic 
Heavy Ion Collisions (URHIC), then a wide variety of probes and observables 
could be used to diagnose and map out its physical properties. 

Measured quenching patterns of pions and $\eta$ mesons~\cite{phenix_pi0} 
already provided a direct evidence for the creation of a strongly interacting
Quark Gluon Plasma (sQGP) in central Au+Au collisions at $\sqrt{s} = 200$ 
$A$GeV~\cite{Gyulassy:2003mc}-\cite{Gyulassy:2004zy}. Further, rare heavy 
quark jets are considered to be excellent independent probes of the 
sQGP~\cite{Brambilla}, because their high mass ($m_c \approx 1.2$ GeV, 
$m_b \approx 4.75$ GeV) changes the sensitivity of the energy loss mechanisms 
in a well defined way~\cite{Dead_cone}-\cite{Armesto:2005iq} relative to those 
of light quark and gluon jets~\cite{Gyulassy:2003mc}-\cite{Vitev:2002pf}. 
Another advantage of heavy quarks jet quenching is that gluon jet fragmentation
into heavy mesons can be safely neglected. However, one disadvantage of heavy 
meson tomography is that direct measurements of identified high $p_\perp$ $D$ 
and $B$ mesons are very difficult with current detectors and RHIC 
{luminosities} \cite{Harris:2005gn}. Therefore, the first experimental studies 
of heavy quark attenuation at RHIC have focused on the attenuation of their 
single (non-photonic) electron decay products~\cite{elecPHENIX,elecSTAR}.

The first preliminary data~\cite{elecPHENIX,elecSTAR}
{surprisingly suggest} that single electrons with $p_\perp\sim 5$ GeV may 
experience elliptic flow and suppression patterns similar to light hadrons. 
It was measured that the suppression of non-photonic electrons, which is 
expressed in terms of the nuclear modification factor $R_{AA}^e(p_\perp)= 
dN(AA\rightarrow e)/(N^{bin}_{AA}dN(pp\rightarrow e)) $, reaches a value 
$\sim 0.1-0.4$ at $p_\perp\sim 4-8$ GeV. Significant reduction at high 
$p_\perp$ single electrons suggests sizeable heavy quark energy loss. 

Motivated by these data, in~\cite{DGVW_PLB} we applied the theory of heavy 
quark radiative energy loss~\cite{Dead_cone}-\cite{Armesto:2005iq} to 
predict the quenching pattern of single electrons from the decay of high 
$p_\perp$ open charm and bottom hadrons. We showed that because the heavy 
quark ``dead cone'' effect~\cite{Dead_cone} is large - especially for bottom
quarks - radiative energy loss predictions for $R^e_{AuAu}$ are significantly 
above $0.5$ as long as realistic parameter values are used. Therefore, the 
puzzle raised by the non-photonic single electron data is whether these data 
can be explained by the energy loss mechanisms in QGP? 

This proceeding mainly concentrate on a theoretical overview of the heavy 
quark energy loss puzzle posed by the single electron data. We start the 
proceedings with a brief overview of heavy quark 
production and radiative heavy quark energy loss mechanisms in QGP. We then 
study the bottom contribution to the single electron spectra, and show that 
this contribution is significant, and can not be neglected in the computation 
of single electron suppression. We then compute the single electron suppression
from the radiative energy loss mechanisms, and show that radiative energy loss 
alone leads to a disagreement with the single electron data, as long as 
realistic gluon rapidity density is taken into account. Finally, we 
concentrate on the collisional energy loss, and show that the inclusion 
of this additional mechanism may lead to a better agreement with the single 
electron data.

\section{Single electron suppression from radiative energy loss} 

In this section we will compute the single electron suppression which comes 
from the radiative energy loss mechanisms, and show that the prediction 
significantly underestimate the single electron suppression as long as 
realistic values of gluon rapidity density are used.

\begin{figure}[h]
\vspace*{3.2cm} \includegraphics{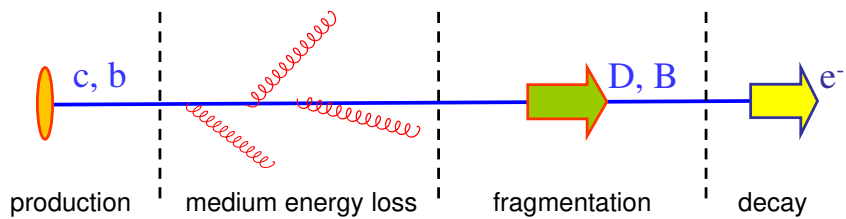}
\caption{A simplified scheme showing how are the non-photonic single electrons 
obtained from QGP.}
\label{scheme}
\end{figure}

To compute the single electron suppression, we start from Fig.~\ref{scheme}, 
which shows the simplified scheme of how the non-photonic 
single electrons are obtained from QGP. We see that, in order to compute the 
single electron spectra, we need to know the initial heavy quark distributions 
from perturbative QCD, heavy flavor energy loss, heavy quark fragmentation 
into heavy hadrons, $H_Q$, and $H_Q$ decay into leptons. The cross section is 
schematically written as (see~\cite{DGVW_PLB,WHDG}): 
\begin{eqnarray}
\frac{E d^3\sigma(e)}{dp^3} &=& \frac{E_i d^3\sigma(Q)}{dp^3_i}
 \otimes
{P(E_i \rightarrow E_f )}
\otimes D(Q \to H_Q) \otimes f(H_Q \to e) \nonumber \; ,
\end{eqnarray}
where $\otimes$ is a generic convolution. The electron decay
spectrum, $f(H_Q \to e)$, includes the branching ratio to electrons.
The change in the initial heavy flavor spectra due to energy loss is 
denoted~$P(E_i \rightarrow E_f)$.

\subsection{Initial heavy quark $p_\perp$ distributions}

One of the main advantages of heavy quarks is their large mass (i.e. 
$M_Q \gg \Lambda_{QCD}$), which, in principle, makes perturbative calculations 
of heavy quark production possible.

An extensive study of perturbative heavy quark $p_\perp$ distributions can be
found in the following papers~\cite{Cacciari:2005rk,MNR} and references 
therein. By using these papers we can perturbatively compute and compare the 
charm and bottom $p_\perp$ distributions. To compute the initial heavy quark 
$p_\perp$ distributions in central rapidity region ($|y|<0.5$) we used the MNR 
code~\cite{MNR}. We assume the same mass and factorization scales as in 
Ref.~\cite{Vogt}, that is we use $M_c = 1.2$~GeV ($M_b = 4.75$~GeV) for charm 
(bottom) mass. For simplicity, we have concentrated only on bare quark 
distributions ($<k^2_\perp >= 0$~GeV$^2$), and the runs were performed by 
using CTEQ5M parton distributions. Initial $p_\perp$ distributions used in our 
computations are shown on the left panel of Fig.~\ref{quark_dist}. From the 
left panel of Fig.~\ref{quark_dist} we see that at low momentum region, bottom 
contribution is negligible compared to charm contribution. On the other hand, 
at higher momentum region these two contributions become approximately the 
same. This is a first indication that bottom contribution may become important 
in the single electron spectrum.

\subsection{Radiative heavy quark energy loss}

There are three medium effects that control heavy quark radiative energy loss. 
These effects are 1)~Ter-Mikayelian, or massive gluon 
effect~\cite{DG_TM,Kampfer}, 2)~Transition radiation~\cite{Zakharov,MD_TR} 
which comes from the fact that medium has finite size and 3)~Medium induced 
radiative energy loss~\cite{Djordjevic:2003zk}, which corresponds to the 
additional gluon radiation induced by the interaction of the jet with the 
medium. In~\cite{DGW_PRL} we showed that first two effects are not important 
for the heavy quark suppression, since they lead to a change of $\pm 0.1$ in 
the charm and bottom $R_{AA}$. We therefore neglect these two effects in this 
proceedings, and concentrate only on the medium induced gluon radiation 
spectrum, given by:

\bea 
\frac{ d N_{ind}^{(1)}}{d x} &=&  \frac{C
_{F}\alpha_{S}}{\pi} 
\frac{L}{\lambda_g} \int_0^\infty 
\frac{ 2 \mathbf{q}^2 \mu^2 d\mathbf{q}^2}{( \frac{4 E x}{L} )^{2} 
+ (\mathbf{q}^{2} + M^{2}x^{2} + m_{g,p}^{2})^{2}} 
 \int \frac{ d\mathbf{k}^2 \; \theta (2 x (1-x) p_{\perp}-
|\mathbf{k}|)} {(( |\mathbf{k}|-|\mathbf{q}|)^{2} + \mu^{2})^{3/2} 
(( |\mathbf{k}|+|\mathbf{q}|)^{2} + \mu^{2})^{3/2}} 
\nonumber \\
&& \hspace*{3cm}\times \left\{ \mu^2+ (\mathbf{k}^{2}-\mathbf{q}^{2}) 
\frac{\mathbf{k}^{2} - M^{2}x^{2} - m_{g,p}^{2}}
{\mathbf{k}^{2} + M^{2}x^{2} + m_{g,p}^{2}} \right\}. 
\label{gloun_rad_1}
\eea
Here, $\mathbf{k}$ is the transverse momentum of the radiated gluon and 
$\mathbf{q}$ is the momentum transfer to the jet. $M$ is heavy quark mass, 
$\mu \approx 0.5$~GeV is Debye mass, $\lambda_g \approx 1$~fm is the mean free 
path, $L \approx 5$~fm is assumed thickness of the medium, 
$m_{g,p}=\mu / \sqrt{2}$~\cite{DG_TM} is gluon mass in the medium, 
$m_{g,v} \approx \Lambda_{QCD}$ is gluon mass in the vacuum and 
$E=\sqrt{p_{\perp}^2+M^2}$ is initial heavy quark energy. We assume constant 
$\alpha_{S}=0.3$ in this study. 

\subsection{Heavy quark $p_\perp$ distributions before and after quenching}

By knowing the initial heavy quark $p_\perp$ distributions and the heavy quark 
radiation spectrum, we are able to compute the heavy flavor $p_\perp$ distributions 
after quenching. For this purpose, we generalized the multigluon fluctuation 
approach from~\cite{GLV_suppress} to the case of finite mass. We here give only
the final results, while for more details the reader should refer 
to~\cite{DGW_PRL,DGVW_PLB,WHDG}. 
 
\begin{figure}[h]
\vspace*{5.0cm} \includegraphics{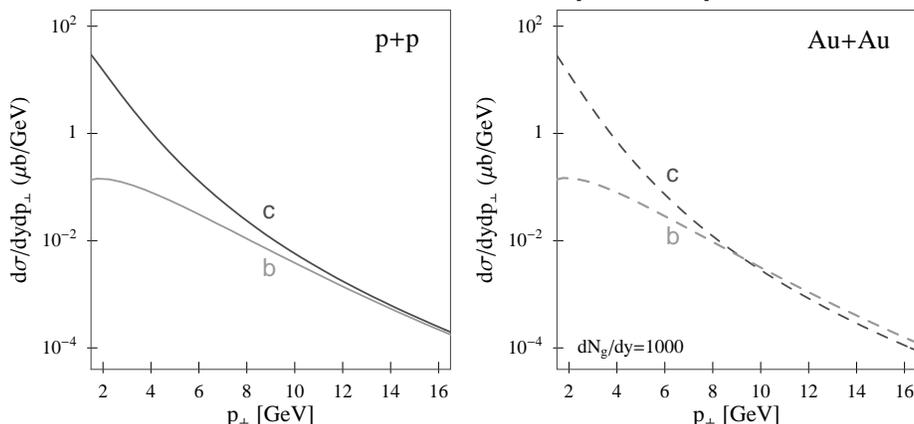}
\caption{ Heavy quark $p_\perp$ distributions before and after quenching are 
shown on the left and right panels respectively. For the right panel, assumed 
gluon rapidity density is $dN_g/dy=1000$~\cite{Gyulassy:2004zy,PHOBOS}. Dark (light) gray curves correspond 
to charm (bottom) quarks.}
\label{quark_dist}
\end{figure}

Figure~\ref{quark_dist} compare $p_\perp$ distributions for heavy quarks before
and after quenching. We see that, while before quenching ($p+p$ collisions) 
charm $p_\perp$ distribution is always larger than bottom $p_\perp$ 
distribution, after quenching ($Au+Au$ collisions) bottom $p_\perp$ 
distribution starts to dominate the spectra after $\sim 9$~GeV. This is 
expected, having in mind that bottom loose significantly less energy
than charm quark (compare dot-dashed curves in Fig.~\ref{comparison}), and it 
is therefore less suppressed than charm quark.

\subsection{Radiative energy loss prediction for single electron suppression}
 
\begin{figure}[h]
\vspace*{5.1cm} \includegraphics{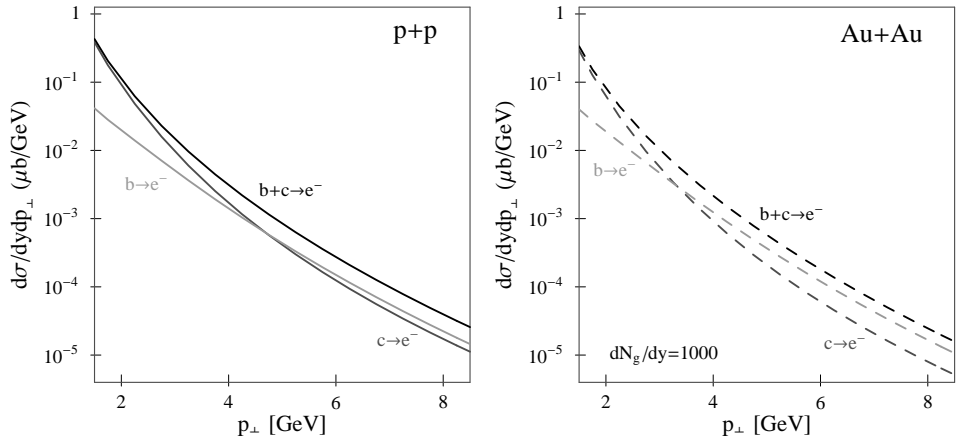}
\caption{ Single electron $p_\perp$ distributions before and after quenching 
are shown on the left and right panels respectively. For the right panel, 
assumed gluon rapidity density is $dN_g/dy=1000$. Dark (light) gray curves 
correspond to charm (bottom) quark contribution to single electrons. Black 
curves show total (charm and bottom) single electron $p_\perp$ distributions.}
\label{electron_dist}
\end{figure}

In this subsection we show the single electron $p_\perp$ distributions before 
and after quenching, which are obtained after fragmentation and decay of heavy 
quark $p_\perp$ distributions from the previous subsection. For more details 
on how we obtained these $p_\perp$ distributions, the reader should refer 
to~\cite{DGVW_PLB,Cacciari:2005rk}. 

Figure~\ref{electron_dist} compare $p_\perp$ distributions for single electrons
before and after quenching. We see that, in the case when quarks are not 
quenched, bottom contribution to single electron spectrum becomes comparable to
charm contribution at $p_\perp\sim 5.5$~GeV. For $dN_g/dy=1000$ case, the 
crossover between charm and bottom contribution is reduced to 
$p_\perp\sim 3.5$~GeV. Therefore, in QGP, electrons in the $p_\perp\sim 5$ GeV 
region have to be sensitive to both $b$ and $c$ quark quenching.

\begin{figure}[h]
\vspace*{5.cm} \includegraphics{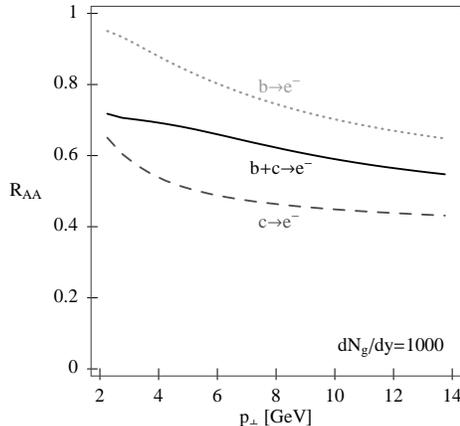}
\caption{Radiative energy loss prediction for single electron suppression. 
Assumed initial gluon rapidity density is $dN_g/dy=1000$. Dashed (dotted) curve
shows what would be the single electron suppression if single electrons would 
have only charm (bottom) contribution. Full curve shows the non-photonic 
single electron suppression by taking into account both charm and bottom 
quark contributions.}
\label{el_supp_rad}
\end{figure}

We can now divide $p_\perp$ distributions before and after quenching to obtain
the single electron suppression that comes from radiative energy loss 
computations (see Fig.~\ref{el_supp_rad}). We see that, for realistic values of
gluon rapidity density, radiative energy loss predicts small single electron 
$R_{AA} \approx 0.7 \pm 0.1$, which is in disagreement with RHIC single 
electron data~\cite{elecPHENIX,elecSTAR}. One possible solution to this problem
is to enhance the gluon rapidity density to a maximal value, which would still
fit the lower boundary on pion suppression data. In~\cite{DGVW_PLB}, we showed 
that for $dN_g/dy \sim 3500$ the non-photonic single electrons can be 
suppressed to $R_{AA}\sim 0.5 \pm 0.1$. However, such large values of gluon 
rapidity density would violate the bulk entropy bounds. Similarly, in 
Ref.~\cite{Armesto:2005iq}, it was found that a similarly excessive transport
coefficient~\cite{Baier:2002tc}, $\hat{q}_{eff}\sim 14$ GeV$^2$/fm, was 
necessary to approach the measured suppression of electrons from charm jet 
decay. This finding raised the question of what is the cause for the observed 
discrepancy between theoretical predictions and experimental results.

\section{Collisional energy loss as a solution to the problem?}

Recent studies~\cite{Mustafa,Dutt-Mazumder} suggested that one of the basic 
assumptions that pQCD collisional energy loss is negligible compared to 
radiative~\cite{Bjorken:1982tu} may be incorrect. 
In~\cite{Mustafa,Dutt-Mazumder} it was shown that, for a range of parameters 
relevant for RHIC, radiative and collisional energy losses for heavy quarks 
were in fact comparable to each other, and therefore collisional energy loss 
can not be neglected in the computation of jet quenching. This result came as 
a surprise because from the earlier 
estimates~\cite{Bjorken:1982tu}-~\cite{Wang:1994fx}, the typical collisional 
energy loss was erroneously considered to be small compared to the radiative 
energy loss. 

However, the computations~\cite{Mustafa}-\cite{Wang:1994fx} were done in an 
infinite QCD medium, while the medium created in URHIC has finite size. A 
recent paper by Peigne {\em et al.}~\cite{Peigne} is the first study that made 
an attempt to include finite size effects in the collisional energy loss. This 
work suggested that collisional energy loss is large only in an ideal infinite 
medium case\footnote{In the case of an infinite QCD medium, the collisional 
energy loss per unit length $dE_{el}/dL$ is computed by assuming that the jet 
is produced at $x_0 = −\infty$. The energy loss for a finite size medium is than
(simplistically) calculated by multiplying this $dE_{el}/dl$ with the thickness 
L of the medium.}, while finite size effects lead to a significant reduction of the 
collisional energy loss. However, this paper did not completely separate 
collisional and radiative energy loss effects. Consequently, it remained 
unclear how important are the finite size effects on the collisional energy 
loss. 
 
\begin{figure}[h]
\vspace*{5.cm} \includegraphics{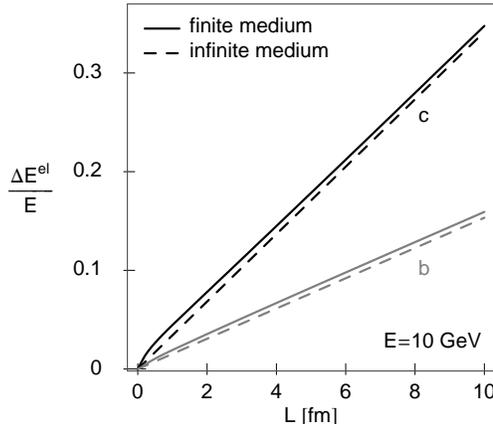}
\caption{Fractional collisional energy loss is shown as a function of 
thickness of the medium for charm and bottom quark jets (upper 
and lower set of curves respectively). Full curves correspond to finite 
medium case, while dash-dotted curves correspond to infinite medium case. 
Initial momentum of the jet is 10 GeV. We assume constant $\alpha_{S}=0.3$.}
\label{finite_size_effects}
\end{figure}

Therefore, it became necessary to consistently compute (only) the collisional 
energy loss in a finite size QCD medium, and see whether the collisional 
energy loss should be taken into account in the computation of jet quenching. 
In~\cite{MD_coll} we provided a detailed study of the $0^{th}$ order 
collisional energy loss in a finite size QCD medium created in URHIC. Contrary 
to~\cite{Peigne}, we find that a finite size medium does not have a large effect 
on the collisional energy loss, as shown in Fig.~\ref{finite_size_effects}.

\begin{figure}[h]
\vspace*{5.cm} \includegraphics{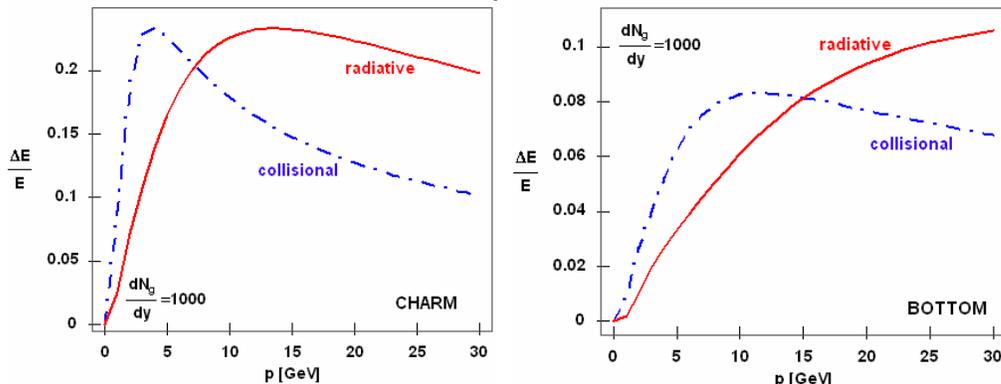}
\caption{Comparison between collisional and medium induced radiative 
fractional energy loss is shown as a function of momentum for charm and bottom 
quark jets (left and right panels respectively). Full curves show the 
collisional energy loss, while dot-dashed curves show the net radiative energy 
loss. Assumed thickness of the medium is $L=5$~fm and $\lambda=1$~fm.}
\label{comparison}
\end{figure}

Comparison between collisional and medium induced radiative energy 
loss~\cite{MD_coll} is shown in Fig.~\ref{comparison}. We see that collisional 
and medium induced radiative energy losses are comparable. Therefore, 
consistently with the claims in Refs.~\cite{Mustafa,Dutt-Mazumder}, we see that 
collisional energy loss is important, and has to be included in the computation 
of jet quenching. 
 
\begin{figure}
\vspace*{5.7cm} \includegraphics{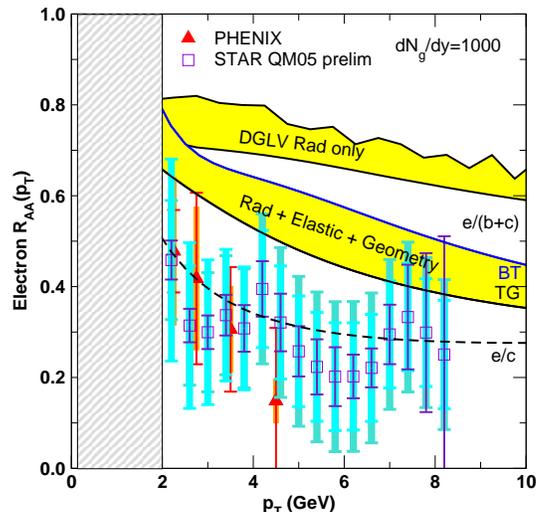}
\caption{The suppression factor, $R_{AA}(p_\perp)$, of non-photonic electrons 
from decay of quenched heavy quark (c+b) jets is compared to 
PHENIX~\cite{elecPHENIX} and preliminary STAR data~\cite{elecSTAR} 
data in central Au+Au reactions at 200~AGeV.  Assumed initial gluon rapidity 
density is $dN_g/dy=1000$. The upper yellow band 
from~\protect{\cite{DGVW_PLB}} takes into account radiative energy 
loss only, using a fixed $L=6$ fm; the lower yellow band includes both 
collisional and radiative energy losses as well as jet path length 
fluctuations~\protect{\cite{WHDG}}. The dashed curve shows the electron 
suppression using radiative and TG~\cite{TG} collisional energy loss with 
bottom quark jets neglected. Figure adapted from~\protect{\cite{WHDG}}.}
\label{electron_supp_CR}
\end{figure}

The results from Fig.~\ref{comparison} are still not included in the 
computation of jet quenching. However, in~\cite{WHDG} we computed the single 
electron suppression by taking into account medium induced radiative energy 
loss together with two different computations (TG~\cite{TG} and BT~\cite{BT}) 
of collisional energy loss done in an infinite QCD medium (see 
Fig.~\ref{electron_supp_CR}). We see that with collisional energy loss, it is
possible to reach single electron $R_{AA} \approx 0.5 \pm 0.1$ even at 
$dN_g/dy = 1000$. While the preliminary data suggest that $R_{AA} < 0.4$, these 
predictions are consistent with the data within present experimental and 
theoretical errors. Therefore, we may conclude that combined collisional and 
radiative pQCD approach may be able to solve a substantial part of the 
non-photonic single electron puzzle. 
 
We note that collisional energy loss in a finite 
size QCD medium falls between the two different computations~\cite{TG,BT} used 
in~\cite{WHDG} (for more details see~\cite{MD_coll}). Therefore, we expect that
the single electron suppression results - computed with finite size collisional
energy loss~\cite{MD_coll} - should be inside the middle yellow region 
presented in Fig.~\ref{electron_supp_CR}.  

We also note that (see~\cite{DGVW_PLB,WHDG}), any proposed solution
of this puzzle must also be consistent with the extensive pion quenching 
data~\cite{phenix_pi0}. In~\cite{WHDG} we show that the simultaneous inclusion 
of path fluctuations together with radiative and collisional energy loss makes 
it possible to satisfy $R_{AA}^e<0.5 \pm 0.1$ without violating the bulk 
$dN_g/dy=1000$ entropy constraint~\cite{Gyulassy:2004zy,PHOBOS} and without violating the pion quenching 
constraint $R_{AA}^{\pi^0}\approx 0.2\pm 0.1$ now observed out to 20 GeV.

Finally, note that in this study we used constant coupling $\alpha_S=0.3$. 
However, the radiative and elastic energy losses depend on the coupling, 
$\Delta E^{rad} \propto \alpha_S^3$ and $∆E^{el} \propto \alpha_S^2$ . 
Future calculations will have to relax the current fixed $\alpha_S$ approximation
and allow it to run. Additionally, the running coupling in collisional 
and radiative processes may be different, since these processes might 
probe different energy scales.

\section{Conclusion}

In this proceeding we applied the theory of heavy quark energy loss to 
non-photonic single electron suppression. We showed that bottom quark 
contribution can not be neglected in the computation of single electron 
spectra. Additionally, we showed that the recent single electron data lead to 
significant discrepancies with theoretical predictions based only on radiative 
energy loss, as long as realistic values of gluon rapidity density are taken 
into account. Finally, we introduced the collisional energy loss mechanisms, 
and showed that combined collisional and radiative pQCD approach may be able 
to solve a substantial part of the non-photonic single electron puzzle.

\subsection*{Acknowledgments} I thank Miklos Gyulassy, Simon Wicks,
William Horowitz and Ramona Vogt for the collaboration on the heavy quark 
energy loss projects. Valuable discussions with Ulrich Heinz are gratefully 
acknowledged. This work is supported by the U.S. Department of Energy, grant 
DE-FG02-01ER41190.

\end{document}